\documentstyle[amsfonts,twocolumn,floats,epsf,prl,aps]{revtex}
\begin{document}
\draft
\wideabs{%
\title{Coarsening Dynamics of Crystalline Thin Films}
\author{Martin Siegert}
\address{Department of Physics, Simon Fraser University,
         Burnaby, British Columbia, Canada V5A 1S6}
\date{\today}
\maketitle
\begin{abstract}
The formation of pyramid-like structures in thin-film growth on substrates
with a quadratic symmetry, e.g., $\{001\}$ surfaces, is shown to exhibit
anisotropic scaling as there exist two length scales with {\it different\/}
time dependences. Analytical and numerical results indicate that for most
realizations coarsening of mounds is described by an exponent $n=1/(3\sqrt2)$.
However, depending on material parameters, $n$ may lie between 0 (logarithmic
coarsening) and 1/3. In contrast, growth on substrates with triangular
symmetries ($\{111\}$ surfaces) is dominated by a single length
$\sim t^{1/3}$.
\end{abstract}
}
\narrowtext
The formation of pyramids or mounds in growth processes such as molecular
beam epitaxy has been observed in numerous
experiments\cite{experiments,Johnson,Stroscio}
and computer simulations\cite{Johnson,Stroscio,SP94,computer_simulations,Amar}.
This three-dimensional growth mode is caused by step-edge barriers that
hinder the diffusion of adatoms across step edges. This generates a surface
diffusion current {\bf j} that causes the initial high-symmetry orientation of
the surface to become unstable\cite{Villain,Johnson}. As a consequence the
surface breaks up forming pyramid-like mounds, the side orientations of which
correspond to stable zeros of the diffusion current\cite{SP94,KPS}. At this
level this phenomenon is theoretically well understood. Conversely,
the observation that the evolving surface morphology coarsens as more and more
material is deposited, is poorly understood. The similarity with ordering
phenomena\cite{S_cam} like spinodal decomposition has led to the assumption
that correlation functions like the height-height correlation function
\begin{equation}
C({\bf r},t)=L^{-2}\sum_{\bf x}\langle h({\bf r}+{\bf x},t)h({\bf x},t)\rangle
\label{corrfct}
\end{equation}
or its Fourier transform, the structure factor
\begin{equation}
S({\bf k},t)=\langle\widehat h({\bf k},t)\widehat h({\bf -k},t)\rangle\ ,
\label{structfac}
\end{equation}
depend only on a single length scale, the pyramid size $R(t)$, which in turn,
because of the scale invariance of the correlation functions, must have a 
power-law dependence $R(t)\sim t^n$. Here, the surface height $h({\bf r},t)$
is measured in a comoving frame of reference such that the average height is
zero, $\widehat h({\bf k},t)$ is the Fourier transform of $h({\bf r},t)$, and
$L^2$ is the size of the substrate. In numerical simulations and
several experiments the coarsening exponent was found to be close to
$n\simeq1/4$. However, theories that attempt to calculate this exponent are
so far not very convincing.
\par
Surprisingly, the initial assumption that there is only a single length scale
and that the correlation function therefore follow simple scaling laws, was
never really questioned. Only recently\cite{SPZ} it was found, that this
kind of surface dynamics is in some respects different from ordering dynamics
like spinodal decomposition: The dynamical exponent $z$ that describes the
relaxation of shape fluctuations of pyramids is not equal to
the inverse $1/n$ of the coarsening exponent indicating that the dynamical
evolution in these two cases is governed by different length scales. In this
article it will be shown that the coarsening dynamics itself depend on
at least two length scales with {\it different time dependences\/}. Therefore,
the standard scaling assumption for the correlation functions is invalid.
\par
At not too high temperatures, so that desorption can be neglected, the
evolution of the surface in molecular beam epitaxy is described by the
continuity equation
\begin{equation}
\partial_t h = -\Delta\Delta h-\nabla\cdot{\bf j}(\bf m)+\eta\ ,
\label{eq_of_motion}
\end{equation}
where ${\bf m}=\nabla h$ is the slope of the surface profile and $\eta$
represents shot noise due to fluctuations in the deposition flux.
The form of the surface current {\bf j} must be chosen such that it 
describes the above mentioned instability and leads to slope selection.
The form ${\bf j}\sim{\bf m}(1-{\bf m}^2)$ has been 
used\cite{Stroscio,RK}, however, this expression has stable zeros for all
slopes with $|{\bf m}|=1$ regardless of the direction of {\bf m}. Such an
azimuthal symmetry is unrealistic for crystalline films.
The simplest form that describes growth on substrates with a quadratic
symmetry is a current with components
\begin{equation}
j_x=m_x(1-m_x^2-bm_y^2)\,,\ j_y=m_y(1-m_y^2-bm_x^2)
\label{j_quad}
\end{equation}
which leads to a buildup of pyramids with selected slopes $(\pm1,\pm1)/(1+b)$
for $-1<b<1$. We will concentrate on the case $b=0$ first. The relevance
of the parameter $b$ and triangular symmetries applicable to (111) substrates
will be discussed further below.
\par
Coarsening in phase ordering dynamics, e.g., spinodal decomposition or
Ostwald ripening, is a deterministic process, i.e., noise is
irrelevant\cite{Bray}. In the same spirit only the deterministic aspects of
the  dynamics described by Eq.~(\ref{eq_of_motion}) with $\eta=0$ will be
studied here. Effects of the noise are discussed at the end of the article.
\par
\begin{figure}[bt]
\hbox to \columnwidth{\hfil%
\vbox{\hsize 52mm\epsfxsize=52mm\epsfbox{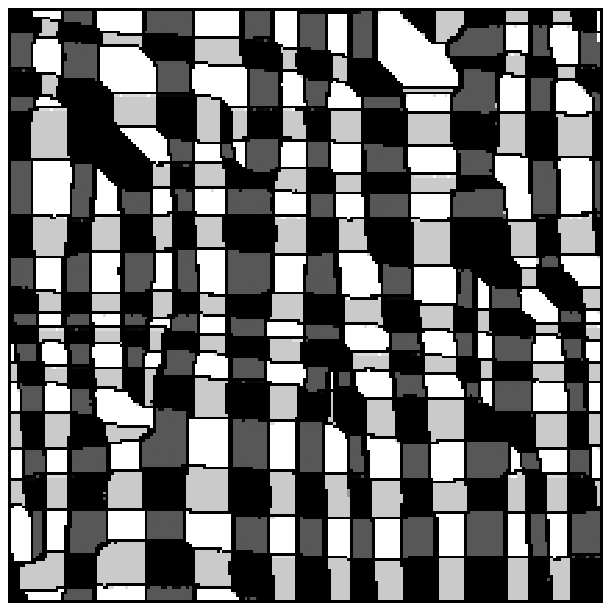}%
a)\hskip 7mm$\uparrow$\hskip 29mm$\uparrow$}
\hfil\vbox to 55mm{\hsize=20mm\vfil%
\epsfxsize=20mm\epsfbox{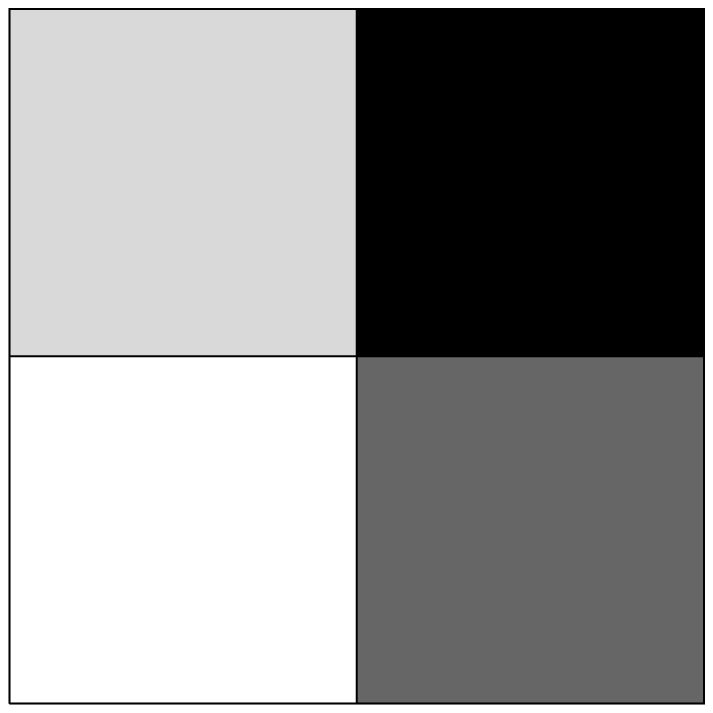}b)\vfil
\epsfxsize=20mm\epsfbox{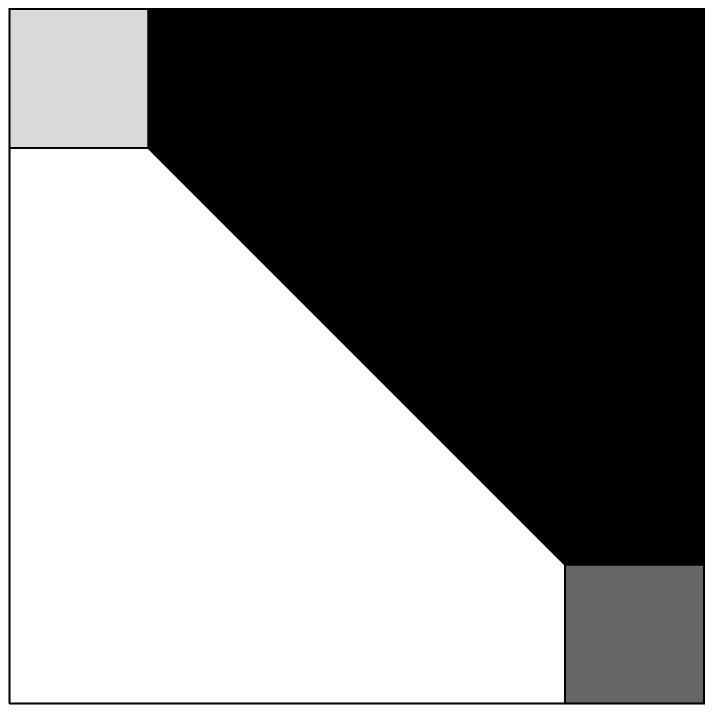}c)\vfil}\hfil}
\caption{a) Domain wall structure obtained from a numerical solution of
Eqs.~(\protect\ref{eq_of_motion}), (\protect\ref{j_quad}); b, c) types of
domain walls: b) pyramid edges, c) roof top. The grey\-scales correspond to
average slopes in the following way: white\,$\equiv$\,$(+1,+1)$,
black\,$\equiv$\,$(-1,-1)$, light grey\,$\equiv$\,$(+1,-1)$,
dark grey\,$\equiv$\,$(-1,+1)$.}
\label{fig:domains}
\end{figure}

The important differences between the problem studied here and phase ordering
dynamics described by the Cahn-Hilliard equation\cite{CH} becomes apparent,
when the domain configurations are plotted as in Fig.~\ref{fig:domains}.
A domain in this context is an area of constant slope corresponding
to one of the four values $(\pm1,\pm1)$. The analogous case in phase
ordering dynamics is described by a four-state clock model\cite{S_Physica}.
However, in that case domain walls do not have any particular orientation,
whereas here domain walls are intersections of planes of constant slopes
and therefore form {\it straight\/} lines. Furthermore, there are two types
of domain walls: Domain walls at which only one component of the slope
changes are aligned along the $x$- and $y$-axes. These are the edges of the
pyramids. Domain walls at which both components of the slope change run at
$45^\circ$ with respect to the principal axes. These latter domain walls
form roof tops as illustrated in Fig.~\ref{fig:domains}c. Domain walls in
systems that phase order give rise to a power-law tail in the structure-factor
called Porod's law\cite{Porod}. On scales $\xi\ll x\ll R(t)$ the height
function close to an edge of a pyramid running in the $y$-direction has the
form $h(x,y)=|x|+y$. Here, $\xi$ is the width of such a  domain wall
[$\xi=\sqrt2$ for (\ref{j_quad})and $b=0$]. Hence, the slope
${\bf m}=({\rm sign}(x),1)$ leads to a singular contribution in the
slope-slope correlation function\cite{slslcor},
$C_{\rm{\bf m},sing}=-2|x|\rho_{\langle100\rangle}$,
where $\rho_{\langle100\rangle}$ is the density of these domain
walls\cite{Bray_rev}. More generally, domain walls that make an angle
$\psi$ with the $y$-axis yield a contribution
$S_{\bf m}=4k^{-3}\delta(\varphi-\psi)\rho_\psi(t)$ to the slope structure
factor, where $\rho_\psi(t)$ is again the density of the domain walls
and $\varphi$ the azimuthal angle such that
${\bf k}=(k\cos\varphi,k\sin\varphi)$. Consequently, the structure
factor~(\ref{structfac}) for crystalline films with a quadratic anisotropy
has a highly anisotropic tail: for $1/R(t)\ll k\ll1/\xi$
\begin{eqnarray}
S({\bf k},t)={4\over k^5}\sum_{\nu=0}^3&%
\Bigl[&\rho_{\langle100\rangle}(t)\delta\left(\varphi-{\pi\over2}\nu\right)
\nonumber\\
     &&+2\rho_{\langle110\rangle}(t)
        \delta\left(\varphi-{\pi\over4}-{\pi\over2}\nu\right)\Bigr]\ ,
\label{skttail}
\end{eqnarray}
where the additional factor 2 in front of the density
$\rho_{\langle110\rangle}(t)$ of domain walls at $45^\circ$ directions
(roof tops) comes from the fact that both components of the slope change at
such a domain wall. Eq.~(\ref{skttail}) is one of the central results of this
article: It shows that the structure factor depends on two different length
scales $1/\rho_{\langle100\rangle}$ and $1/\rho_{\langle110\rangle}$,
specifying the average separation of edges of pyramids and roof tops,
respectively. Whereas for $\rho_{\langle100\rangle}\gg\rho_{\langle110\rangle}$
the former is directly related to the average pyramid size $R(t)$, no
obvious relation exists between $\rho_{\langle110\rangle}$ and $R(t)$.
In fact, {\it there is no reason why $1/\rho_{\langle100\rangle}$ and
$1/\rho_{\langle110\rangle}$ should have the same time-dependence or follow
the same power law}. Therefore, the structure factor does not obey a
scaling law $S({\bf k},t)=R^2(t)s(kR(t))$, not even in a modified
anisotropic form\cite{Rutenberg_aniso}. Hence, all theoretical approaches
that assume such a dependence on a single length scale, are unfounded.
\par
The result that the structure factor is nonzero
only along the ${\langle100\rangle}$ and ${\langle110\rangle}$ directions
as indicated by the delta functions in (\ref{skttail}) is a consequence of
the assumption that the singular part of $C({\bf r},t)$ dominates the
contributions to the tail of $S({\bf k},t)$. For directions other than
these high symmetry directions no singularities exist and one should
expect an exponential decay. Furthermore, although fluctuations of the
domain wall directions are finite\cite{SPZ}, such fluctuations will remove
the delta singularities and replace them by a sharply peaked function with
finite width. Nevertheless, the prediction that $S({\bf k},t)\sim k^{-5}$
in directions normal to the domain walls, and that $S({\bf k},t)$ decays
faster for all other directions is confirmed in numerical
simulations\cite{S_unpubl} and it should also be possible to observe this
behaviour experimentally.
\par
\begin{figure}[bt]
\centerline{\epsfxsize=70mm\epsfbox{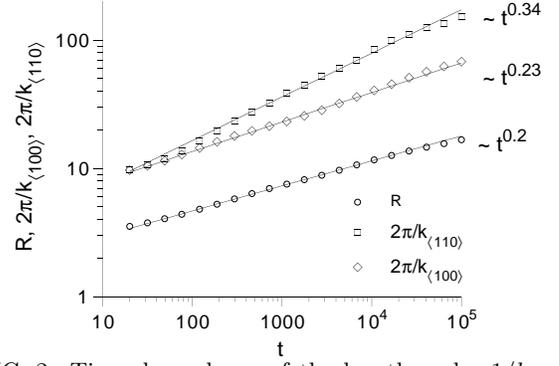}}
\caption{Time dependence of the length scales $1/k_{\langle100\rangle}(t)$ and
$1/k_{\langle110\rangle}(t)$ as defined in
Eqs.~(\protect\ref{k100}), (\protect\ref{k110}). The average pyramid size
$R(t)$ measured as the first zero of the correlation function $C({\bf r},t)$
in the $\langle110\rangle$ directions is plotted as well.}
\label{fig:lengthscales}
\end{figure}

 
Length scales that show the anisotropic scaling of the structure factor
in a numerical solution of Eqs.~(\ref{eq_of_motion},\ref{j_quad}) can be
defined as inverses of moments of the structure factor 
\begin{eqnarray}
k_{\langle100\rangle}(t)&=&\sum\nolimits_{k_x}k_xS((k_x,0),t)\Bigl/
\sum\nolimits_{k_x}S((k_x,0),t)\ ,\label{k100}\\
k_{\langle110\rangle}(t)&=&\sum\nolimits_k\sqrt2kS((k,k),t)\Bigl/
\sum\nolimits_kS((k,k),t)\label{k110}
\end{eqnarray}
in the two directions in question.
Fig.~\ref{fig:lengthscales} shows clearly that the scaling
behaviour of $S({\bf k},t)$ depends on the direction of {\bf k}:
whereas $1/k_{\langle110\rangle}(t)$ is well described by a power law 
$\sim t^{1/3}$, $1/k_{\langle100\rangle}(t)$ grows more slowly.
In fact, in a numerical simulation coarsening stops as soon as the average
distance $1/k_{\langle110\rangle}$ between two roof tops is of the same order
as the system size. {\it At the same time the average pyramid size can be
several orders of magnitude smaller}. The coarsening of pyramids
is enslaved to dynamics of roof tops: A configuration without
any roof tops does not coarsen, i.e., it is a metastable state\cite{S_unpubl}.
Roof tops represent defects in such perfect pyramid lattices and
coarsening proceeds by eliminating such defects. However, because of the more
complicated scaling behaviour indicated in (\ref{skttail}) the theory of Bray
and Rutenberg\cite{BR} can no longer be used to calculate the growth law.
In the limit where the average roof-top distance
$D\sim1/\rho_{\langle110\rangle}$ is much larger than
$1/\rho_{\langle100\rangle}\sim R$ there exists a simple geometric relation
between the two length scales: Coarsening proceeds by elimination of 
finger-like domain configurations, two of which are indicated by arrows in
Fig.~\ref{fig:domains}a. Such fingers are always terminated by roof tops
of length $\sqrt2R(t)$ at either end. If such a finger disappears
within a time interval $\Delta t$, the roof top density
changes by $\Delta(1/D)=-2\sqrt2R/L^2$, where $L$ is some unit of length. In
the same time the average density of pyramid edges changes by
$\Delta(1/R)=-2D/L^2$. It follows that $D\Delta(1/D)=\sqrt2R\Delta(1/R)$ or
$D(t)^{1/\sqrt2}\sim R(t)$. Using the result $D\sim t^{1/3}$ (see below)
one finds $R\sim t^n$ with $n=1/(3\sqrt2)\simeq0.2357$, hardly
distinguishable from 1/4 in a numerical simulation.
\par
\begin{figure}[bt]
\hbox to \columnwidth{%
\vbox{\hsize=16mm\epsfxsize=16mm\epsfbox{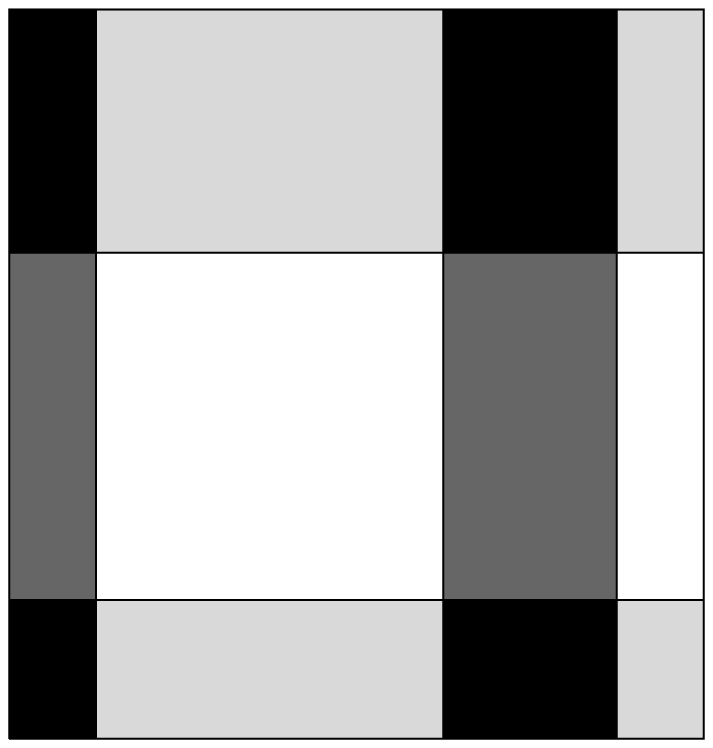}a)}\hfil
\vbox{\hsize=16mm\epsfxsize=16mm\epsfbox{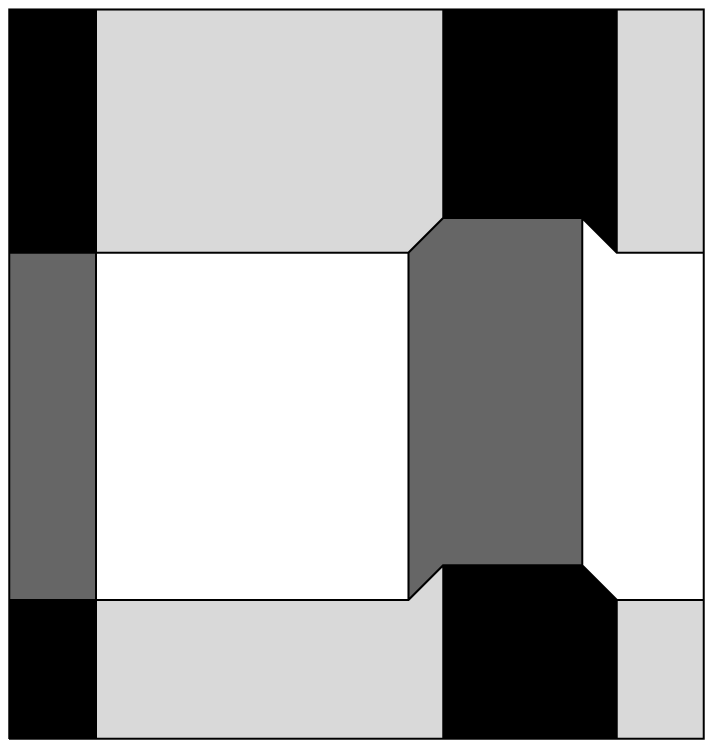}b)}\hfil
\vbox{\hsize=16mm\epsfxsize=16mm\epsfbox{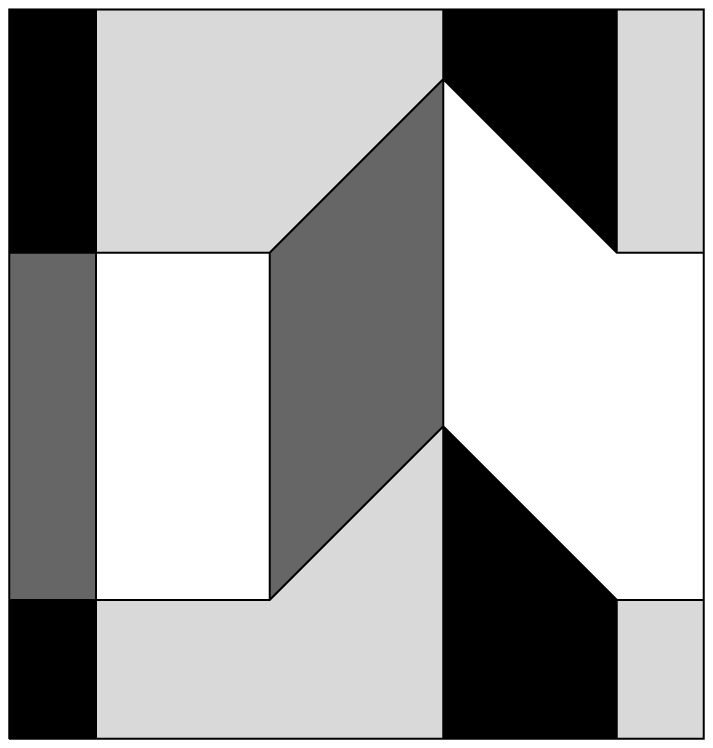}c)}\hfil
\vbox{\hsize=16mm\epsfxsize=16mm\epsfbox{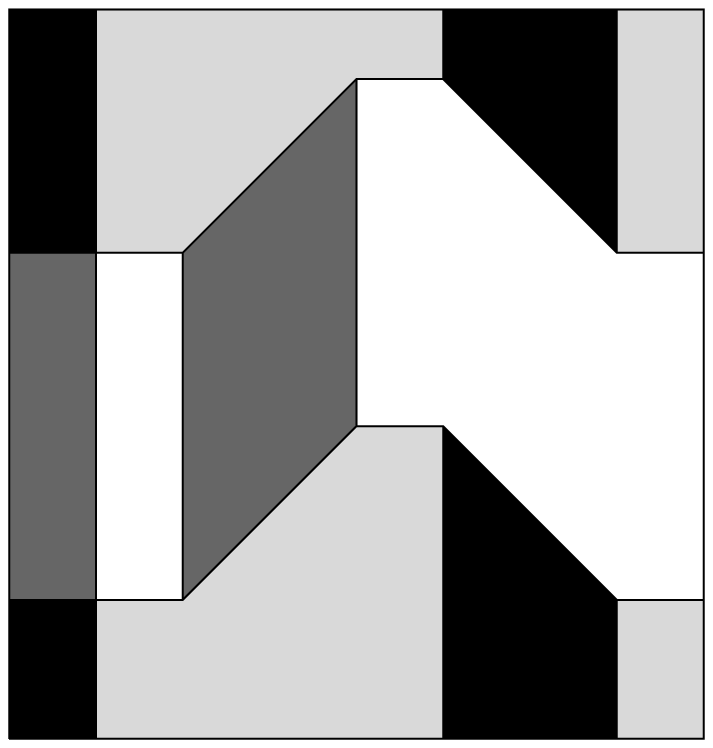}d)}\hfil
\vbox{\hsize=16mm\epsfxsize=16mm\epsfbox{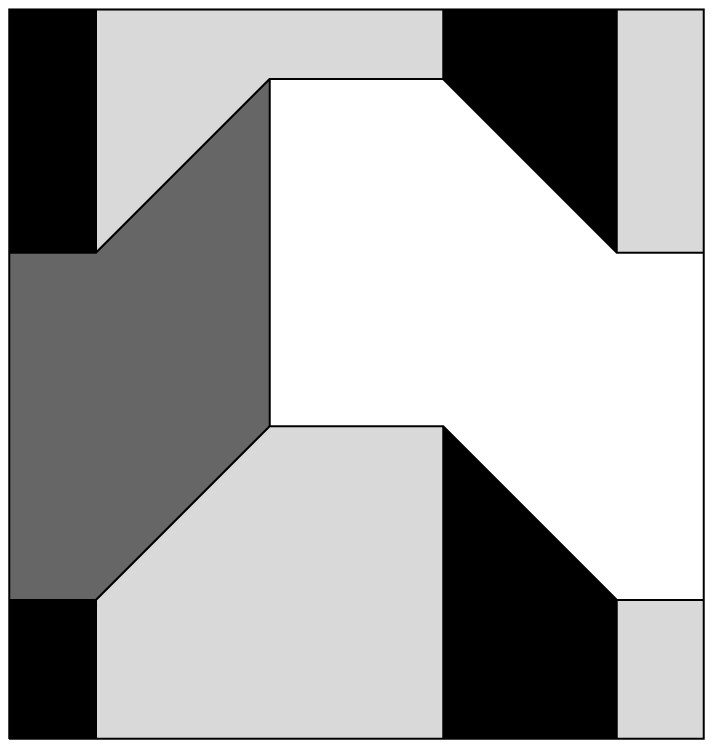}e)}}
\vskip 2mm
\centerline{\vbox to 9mm{\hsize 6mm f)\vfil}%
\epsfxsize=64mm\epsfbox{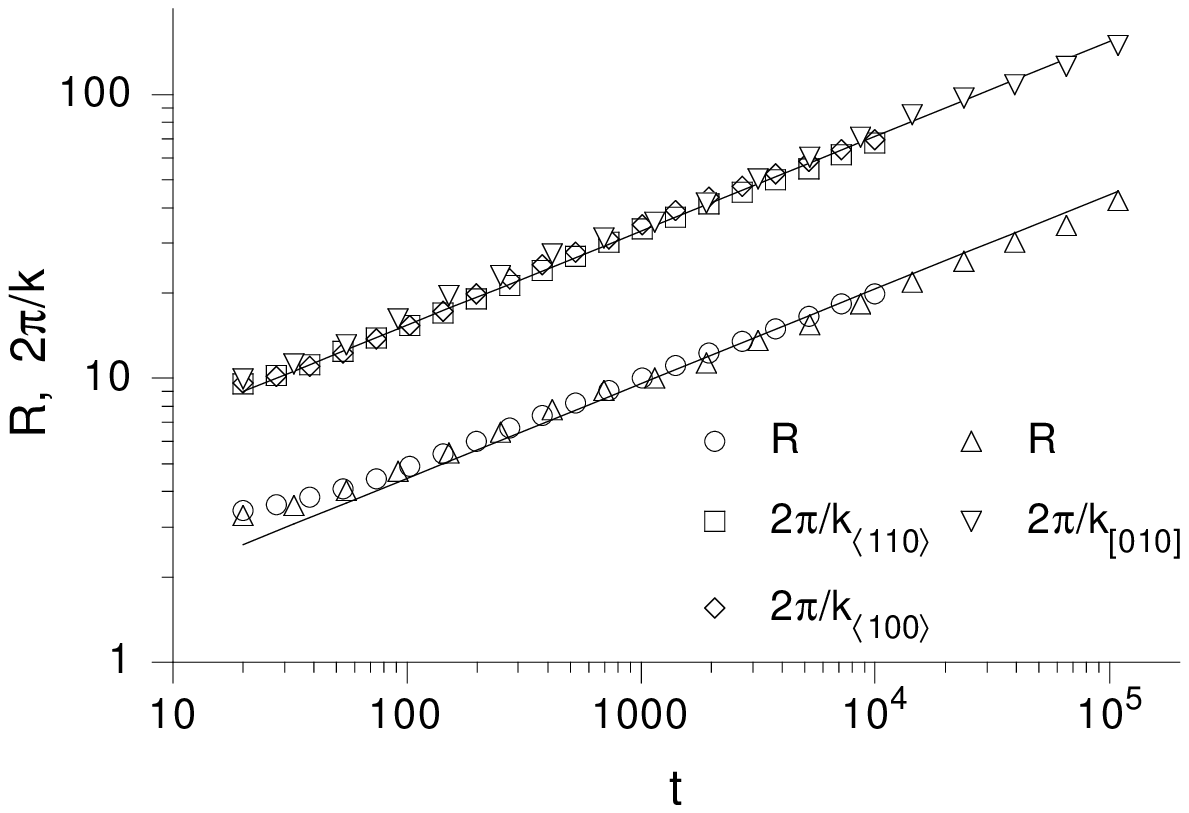}}
\caption{Domain wall motion for films with a quadratic anisotropy:
a-e) schematics of domain wall motions through the introduction of
roof tops, f) coarsening dynamics for the special degenerate case $b=-3/4$.
Data for triangular anisotropies (triangles) are included as well. The
straight lines correspond to power laws $\sim t^{1/3}$.}
\label{fig:dom_coarsen}
\end{figure}

During the temporal evolution of the surface morphology shown in
Fig.~\ref{fig:domains} many more pyramids than roof tops are formed. 
The reason lies in the value $b=0$ that was used in that simulation.
The parameter $b$ describes how the $x$-component of the current depends on
the $y$-component of the slope. Such transverse
currents are typical for effects like edge diffusion, the importance of which
for growth morphologies has been emphasized by Bartelt and Evans\cite{Evans}
and was recently confirmed in an experiment\cite{Poelsema}. It is possible to
assign a ``surface tension'' $\sigma$ to domain walls by
integrating over the domain wall profile,
\begin{equation}
\sigma=\int_{-\infty}^\infty dr_\perp
\left[\left({\partial m_x\over\partial r_\perp}\right)^2
      +\left({\partial m_y\over\partial r_\perp}\right)^2\right]\ ,
\label{sigma}
\end{equation}
where the coordinate $r_\perp$ runs perpendicular to the domain wall and the
domains are assumed to extend to infinity on both sides of the wall.
Using the steady-state solutions for the domain wall profiles that correspond
to Eqs.~(\ref{eq_of_motion}), (\ref{j_quad}), it is found that
$\sigma_{\rm pe}=4/[3\sqrt2(1+b)]$ for pyramid edges and
$\sigma_{\rm rt}=4/[3\sqrt{1+b}]$ for roof tops. Therefore, for $b=0$
the formation of roof tops is suppressed. To understand the coarsening
mechanism it is instructive to study perturbations to a perfect lattice of
pyramids (see Fig.~\ref{fig:dom_coarsen}a-e). 
The movement of domain walls necessarily involves the nucleation of roof tops,
see Fig.~\ref{fig:dom_coarsen}b, c. Such a perturbation does not cost anything,
if $\sqrt2\sigma_{\rm rt}=\sigma_{\rm pe}$, or, if $b=-3/4$. For this value
of the parameter $b$ edge-like domain walls can transform without cost
into roof-top-like domain walls and coarsening is no longer enslaved to the
dynamics of the roof tops. This is convincingly confirmed by numerical
solutions of Eq.~(\ref{eq_of_motion}) with a current (\ref{j_quad}) and
$b=-3/4$. As seen in Fig.~\ref{fig:dom_coarsen}c coarsening is much faster
as all length scales increase as $t^{1/3}$, i.e., {\it the
coarsening of the pyramid size $R(t)$ depends on the parameter $b$\/} that in
turn is determined by microscopic processes like edge diffusion. 
\par
The time scale for the elimination of roof tops is set by the time it takes
to get from configuration~\ref{fig:dom_coarsen}c to
configuration~\ref{fig:dom_coarsen}e. This is a simple diffusion process
as the changes indicated in \ref{fig:dom_coarsen}d do not change the
lengths of the domain walls. This behavour and the time dependence
$\sim t^{1/3}$ suggest that an approach similar to the Lifshitz-Slyozov-Wagner
theory\cite{LSW} can be used here. Assuming that the magnitude
of the slope is slightly less than the asymptotic value, i.e.,
$|m_{x,y}({\bf r},t)|=m_0-\epsilon_{x,y}({\bf r},t)$ with
$\epsilon_{x,y}\ll m_0$ it can be shown that there exists a diffusion
current from the small domains to the larger domains that leads to a coarsening
law $\sim t^{1/3}$. However, the same theory also predicts that the
average deviation $\langle\overline\epsilon(t)\rangle$ should decrease
as $t^{-1/3}$. This latter behaviour is not confirmed in numerical solutions
of Eq.~(\ref{eq_of_motion}) for triangular symmetries or quadratic symmetries
with $b=-3/4$\cite{S_unpubl}. This problem requires further investigation.
\par
In the nondegenerate case with $b>-3/4$ the configurations
\ref{fig:dom_coarsen}b, c are highly suppressed as they correspond to
activated processes. Hence, coarsening
proceeds by eliminating roof tops that were nucleated in the initial stages
of the instability. Consequently, the roof-top distance becomes much larger
than the pyramid size $R$.
\par
Although coarsening exponents close to 1/3 were found in a numerical
simulation\cite{Amar}, the situation that the surface tensions obey the
relation $\sqrt2\sigma_{\rm rt}=\sigma_{\rm pe}$ exactly is unlikely to be
realized in nature. The fact that many experiments measured a coarsening
exponent close to 1/4 instead indicates that typically the pyramid coarsening
is enslaved to the dynamics of the roof tops in which case the theoretical
value of the exponent is $1/(3\sqrt2)$ as explained above. However, this is
no longer correct for thin films with a triangular lattice anisotropy, i.e.,
for (111) substrates. In that case there is only one type of domain wall;
thus one expects fast coarsening in any case. The appropriate form of the
surface current in that case was given in Ref.~\cite{S_Physica}. Results of
a numerical solutions of Eq.~(\ref{eq_of_motion}) are included in 
Fig.~\ref{fig:dom_coarsen}f (triangles). They indeed show that the
pyramid size $R(t)\sim t^{1/3}$. The author is aware of a
single experiment that studied coarsening on a (111) substrate\cite{Clarke}.
The results of that experiment are in perfect agreement with the theory
presented here. Nevertheless, it must be emphasized that even in those cases
where coarsening is $\sim t^{1/3}$ the correlation functions (\ref{corrfct})
and (\ref{structfac}) do not obey a simple scaling law. The result
(\ref{skttail}) that the structure factor scales differently in directions
perpendicular to the domain walls than in all other directions is still valid.
\par
Most of the results derived in this article do not depend on the specific
form of the surface current, e.g., the anisotropic tail of the structure
factor (\ref{structfac}) and the growth exponents follow from the constraints
imposed by the crystalline anisotropies of the growing film, which in turn
severely restrict the possible domain configurations. Only the
results for $\sigma_{\rm pe}$ and $\sigma_{\rm rt}$ are specific to the form
of the surface current (\ref{j_quad}). Because of the existence of several
length scales and the fact that the correlation functions do not obey a simple
scaling law, the method by which the characteristic length scales are measured
becomes important: Whereas the average roof-top distance could be separated
from the average pyramid size using moments of the structure factor in
different directions (\ref{k100}), (\ref{k110}), this cannot be
accomplished in real space, e.g., the height-height correlation function
(\ref{corrfct}) in the $\langle100\rangle$ directions changes its functional
form with time and does not permit the determination of any length scale.
\par
As mentioned in the introduction, most experiments measured a coarsening
exponent for the pyramid size on substrates with quadratic symmetries of
$n\lesssim1/4$ indicating that real systems are nondegenerate, i.e., they
correspond to $b>-3/4$ in Eq.~(\ref{j_quad}). In these cases the roof-top
distance $D$ becomes much larger than the pyramid size $R$ and
$R\sim t^{1/(3\sqrt2)}$ whereas $D\sim t^{1/3}$. However, if the late stages
of the coarsening process are studied on scales smaller than $D$, the pyramid
size coarsens only logarithmically as is typical for activated processes and
as has in fact been predicted for these kind of systems\cite{Shore}.
The measurement of the coarsening behaviour in these cases is severely
complicated by a wide cross-over regime\cite{S_unpubl} where almost any
value between $n=0$ and $n\lesssim1/4$ can be observed. The interpretation
is further complicated when effects of the noise $\eta\ne0$ is included: 
For some time such stochastic fluctuations provide a mechanism to
overcome the activation barriers that suppress the formation of roof tops.
However, the activation barriers are proportional to the pyramid size. Hence,
in the asymptotic regime, where the pyramid size is sufficiently large,
stochastic effects become unimportant. This is in agreement with arguments
presented by Shore {\it et al.}\cite{Shore}. For deposition on
(111) substrates noise is clearly irrelevant.
\par
The author thanks M.~Plischke for many helpful conversations.
This research was supported by the NSERC of Canada.

\end{document}